\documentclass[reprint,aps,prd]{revtex4-1}

\usepackage{amsmath}
\allowdisplaybreaks[4]
\usepackage{multirow}
\usepackage{amsfonts}
\usepackage{xcolor}
\usepackage{graphicx}
\usepackage{siunitx}
\usepackage{float}
\usepackage[
    pdfauthor={Teng Zhang},
    pdftitle={Effects of static and dynamic higher-order optical modes in balanced homodyne readout for future gravitational waves detectors},
    colorlinks=true,
    linkcolor=black,
    urlcolor=blue,
    citecolor=blue
]{hyperref}

\newcommand{\op}[1]{\boldsymbol{#1}}

\newcommand{\TEM}[1]{\ensuremath{\mathrm{TEM}_{#1}}}

\DeclareSIUnit{\sqrthz}{\ensuremath{\sqrt{\text{\hertz}}}}

\begin{document}
\allowdisplaybreaks[4]
\title{Effects of static and dynamic higher-order optical modes in balanced homodyne readout for future gravitational waves detectors}

\author{Teng Zhang$^1$}
\author{Stefan L Danilishin$^1$}
\author{Sebastian Steinlechner$^{1,2}$}
\author{Bryan W Barr$^1$}
\author{Angus S Bell$^1$}
\author{Peter Dupej$^1$}
\author{Christian Gr\"af$^1$}
\author{Jan-Simon Hennig$^1$}
\author{E Alasdair Houston$^1$}
\author{Sabina H Huttner$^1$}
\author{Sean S Leavey$^1$}
\author{Daniela Pascucci$^1$}
\author{Borja Sorazu$^1$}
\author{Andrew Spencer$^1$}
\author{Jennifer Wright$^1$}
\author{Kenneth A Strain$^1$}
\author{Stefan Hild$^1$}
\affiliation{$1$ SUPA, School of Physics and Astronomy, The University 
of Glasgow, Glasgow, G12\,8QQ, UK}
\affiliation{$2$  Institut f\"ur Laserphysik und Zentrum f\"ur Optische Quantentechnologien der Universit\"at Hamburg, Luruper Chaussee 149, 22761 Hamburg, Germany}

\begin{abstract}
With the recent detection  of Gravitational waves (GW), marking the start of the new field of GW astronomy, the push for building more sensitive laser-interferometric gravitational wave detectors (GWD) has never been stronger. Balanced homodyne detection (BHD) allows for a quantum noise (QN) limited readout of arbitrary light field quadratures, and has therefore been suggested as a vital building block for upgrades to Advanced LIGO and third generation observatories.  In terms of the practical implementation of BHD, we develop a full framework for analyzing the static optical high order modes (HOMs) occurring in the BHD paths related to the misalignment or mode matching at the input and output ports of the laser interferometer. 
We find the effects of HOMs on the quantum noise limited sensitivity is independent of the actual interferometer configuration, \textit{e.g.}  Michelson and Sagnac interferometers are effected in the same way. 
We show that misalignment of the output ports of the interferometer (output misalignment) only effects the high frequency part of the quantum noise limited sensitivity (detection noise). However, at low frequencies, HOMs reduce the interferometer response and the radiation pressure noise (back action noise) by the same amount and hence the quantum noise limited sensitivity is not negatively effected in that frequency range. We show that the misalignment of laser into the interferometer (input misalignment) produces the same effect as output  misalignment and additionally  decreases the power inside the interferometer. We also analyze dynamic HOM effects, such as beam jitter created by the suspended mirrors of the BHD. Our analyses can be directly applied to any BHD implementation in a future GWD. Moreover, we apply our analytical techniques to the example of the speed meter proof of concept experiment under construction in Glasgow. We find that for our experimental parameters, the performance of our seismic isolation system in the BHD paths is compatible with the design sensitivity of the experiment. 
 
\end{abstract}

% acronym definitions

\maketitle

%%% =========================================================================

\section{Introduction}
\label{sec:introduction}

After a half-century search, the first detection of gravitational waves in 2015 \cite{GW150914} further inspires the worldwide effort to increase the  sensitivity of laser-interferometric GWDs. As the design sensitivity of the second generation detectors is limited by quantum noise over most of the detection frequency band, the development and implementation of novel techniques which reduce or even circumvent quantum noise is a major task within the detector collaborations \cite{chen2011qnd}.
        
Quantum noise originates from the quantum nature of laser light and manifests itself in two ways. \emph{Shot noise}, or sensing noise, dominates at high frequencies, while \emph{radiation pressure noise}, or back-action noise, dominates at low frequency. At each frequency there is an optimal laser power which balances the two noise sources, giving rise to the so-called Standard Quantum Limit (SQL). Using \emph{quantum non-demolition} (QND) techniques \cite{Braginsky1980}, it is in principle possible to achieve sensitivities beyond the SQL \cite{kimble2001,Corbitt2002}. These techniques often require the readout of a specific quadrature of the interferometer output light field, e.g.\ in the variational readout scheme \cite{kimble2001}. BHD allows for arbitrary readout quadratures and therefore naturally offers itself for this task. Another approach to surpass the SQL is the speed meter topology \cite{Braginsky1990}, in which the speed of a test mass is detected instead of its position. In 2003, Y.~Chen pointed out that the Sagnac interferometer topology behaves as a speed meter, and a proof-of-principle experiment is currently being set up in Glasgow \cite{G2014}. As it turns out, there is no suitable carrier field available in the output port of Sagnac interferometers and so an external local oscillator (LO) is required, which is provided by BHD.

Current GWDs employ a  \emph{dc readout} \cite{stefan2009} (sometimes also referred to as a \emph{homodyne readout}), in which a small differential arm-length offset is introduced that leads to some carrier light in the signal port and which serves as the LO for detection with a single photo detector. The local oscillator power needs to be chosen such that the photon shot-noise is well above the electronic noise of the detector. Now that quantum-noise reduction using squeezed light has become a key ingredient of current detectors \cite{Aasi2013,Grote2013}, the requirements for the local oscillator power and the resulting voltages in the photo detector electronics are close to reaching technical limitations as the squeezing strength further improves \cite{grote2016}. Here, the current-subtracting design of BHD helps to bring the requirements down again to manageable levels \cite{Fritschel2014}.

Thus, there is significant interest in applying BHD in GWDs as an enabling technology for further improvements in the quantum-noise limited sensitivity. So far there is surprisingly little experience with BHD in gravitational wave detectors \cite{Fritschel2014,Sebastian2015}, especially with regards to the requirements and difficulties that come with suspended optics, long baselines and highest sensitivities in the few hundred Hertz regime. Here we develop a framework to investigate and define those effects.  In Sec.~\ref{sec:fundamental}, we introduce a general calculation in BHD readout involving the higher-order mode components; in Sec.~\ref{sec:qnbhd}, we consider how HOMs enter the quantum noise picture that describes interfermeters such as GWDs; in Sec.~\ref{sec:couplingc}, we then illustrate how HOMs come about from misalignment and mismatch in BHD;  in Sec.~\ref{sec:MichelsonBHD}, we derive how static misalignment or mismatch affect the quantum noise limited sensitivity of a  Michelson interferometer; in Sec.~\ref{sec:SSMdesign}, we look at the example of the Glasgow SSM experiment to verify the effects on such a QND techniques candidate configuration;  in Sec.~\ref{sec:jitter}, we calculate the dynamic beam jitter noise coupling in BHD readout.

\section{Fundamentals of Balanced-Homodyne Detection with Higher-Order Modes}
\label{sec:fundamental}
\begin{figure}\label{fig:BHD}
\centering
  \includegraphics[width=1\columnwidth]{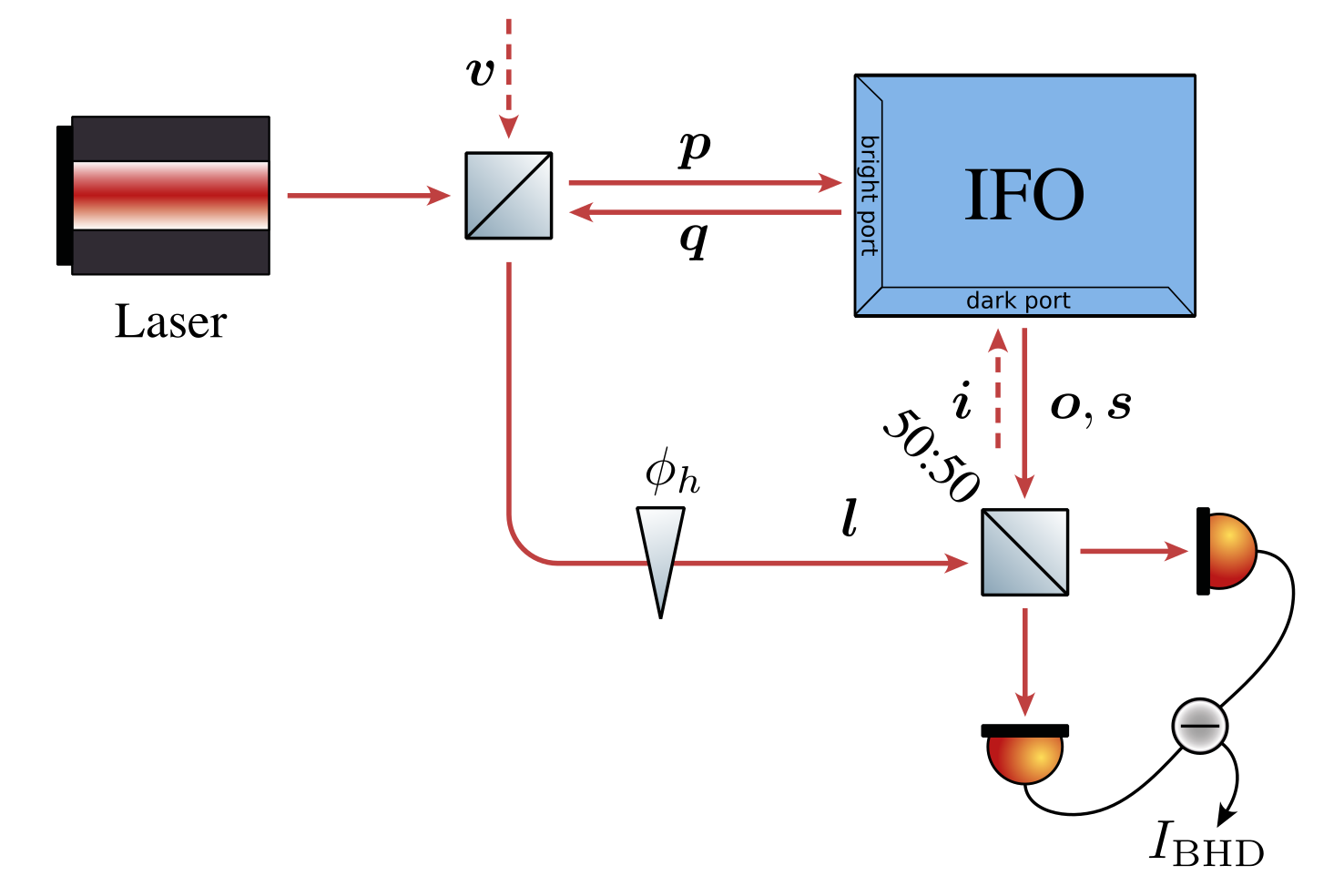}
\caption{Schematic of a balanced homodyne  readout setup of a generic  interferometer.  The input and output field at the bright port of the interferometer are demoted $\op p$ and $\op q$, respectively, while the corresponding fields at the dark port of the interferometer are denoted as  $\op i$ and $\op o$, respectively. Then the $\op o$ field and $\op q$ field enter into the BHD path as signal beam $\op s$ and LO beam $\op l$. There is also a vacuum field $\op v$ that couples into $\op l$ field due to the LO pick off mirror. $\op l$ and $\op s$ are overlapped with each other at the balanced-homodyne beam splitter. The output photo current $I_{\rm BHD}$ is a subtraction of the output of two photodiodes. The homodyne angle $\phi_{h}$ is the relative phase of the two beams entering the BHD.} 
\label{fig:BHD}
\end{figure}
Let us start by establishing the fundamental equations which describe BHD readout with HOMs. We define the time-varying electrical fields of the signal and LO beams as $s(r,t)$ and $l(r,t)$, respectively, where we collected the transverse spatial coordinates in $r$. For both fields, we separate the dc components $S_{mn}$, $L_{mn}$ from the fluctuations $s_{mn}$ and $l_{mn}$, where $m$,$n \geq 0$ are the indices of the Hermite-Gaussian mode expansions, $\TEM{mn}$. A natural reference for the mode expansion is the fundamental mode of the optical instrument, e.g.\ the fundamental mode defined by the arm cavities in a GWD. The two optical fields can then be written as

\begin{equation}
s(r,t)\propto \underset{m,n \geq 0}{\sum} u_{mn}(r,z)[ S_{mn}+ s_{mn}]e^{-i\omega t}+h.c.
\end{equation}
\begin{equation}
l(r,t)\propto \underset{m,n \geq 0}{\sum} u_{mn}(r,z)[ L_{mn}+ l_{mn}]e^{-i\omega t}+h.c.
\end{equation}
where $u_{mn}(r,z)$ is the spatial distribution of the electric field of Hermite-Gaussian modes of orders $m,n$ in the plane transverse of the direction of propagation $z$, $\omega$ is the carrier frequency and $h.c.$ denotes the hermitian conjugate.

Afterwards, the signal and LO beams are overlapped on the BHD beam splitter with a relative phase $\phi_h$ that defines the homodyne angle, i.e.\ the detected light field quadrature. The fields in the two beam splitter outputs are given by
\begin{equation}
P_1=\frac{le^{i \phi_h}+s}{\sqrt{2}},\quad P_2=\frac{-le^{i \phi_h}+s }{\sqrt{2}}.
\end{equation}
These fields are detected by two photodiodes, and the resulting photocurrents are subtracted from each other, resulting in the output photocurrent 
\begin{equation}
\begin{aligned}
I_{\rm BHD}
 &\propto P_1P_1^{\dagger}-P_2P_2^{\dagger}\\
&=
\sum_{m,n \geq 0}
(L_{mn}+l_{mn})(S_{mn}+ s_{mn})^{\dagger}e^{i\phi_h}
+h.c.
\end{aligned}
\label{eq:photocurrent}
\end{equation}
To simplify the notation, in the following we use a single index $j$ to enumerate the mode indices $mn$, i.e.\ for $j=0$, $\{mn\}=\{00\}$; $j=1$, $\{mn\}=\{01\}$; $j=2$, $\{mn\}=\{02\}$, etc.

For future purpose, we separate the BHD photocurrent into the classical dc and the fluctuation parts using so-called \textit{two-photon formalism} \cite{Caves3068,Caves3093}  which is used to describe the  fields using a two-dimensional vector of two orthogonal quadrature amplitudes. Then the dc components and fluctuations in signal beam and LO beam are defined as $\op S, \op s, \op L,\op l$, in which e.g.\ ${\op s}=(\op s_c,\op s_s)^{\rm{T}}$, where the superscript $\rm T$ stands for transpose.  The additional homodyne angle $\phi_h$ is used to single out the particular readout quadrature. Mathematically, it means that the LO field needs to be multiplied by a rotation matrix of the following form:
\begin{equation}
\mathbb{H}_{\phi_h}=
\left[
\begin{array}{cc}
\cos(\phi_h) & -\sin(\phi_h) \\
\sin(\phi_h) & \cos(\phi_h)\\
\end{array}
\right].
\end{equation}
Then the classical dc part reads
\begin{equation}
\label{eq:Iclassic}
I_{\rm BHD}^{\rm dc}\propto \underset{j \geq 0}\sum \op S^{\dagger}_{j}\mathbb{H}_{\phi_h}\op L_{j}+h.c.\,,
\end{equation}
while the fluctuating part, containing classical and quantum noise as well as modulation sidebands, is given by
\begin{equation}
\begin{split}
\label{eq:Ifluctuation}
I_{\rm BHD}^{\rm fl}\propto \underset{j \geq 0}{\sum}   \op s^{\dagger}_{j}\mathbb{H}_{\phi_h}\op L_{j}\\
+\underset{j \geq 0}\sum \op S^{\dagger}_{j}    \mathbb{H}_{\phi_h}\op l_{j}
+\underset{j \geq 0}\sum   \op s^{\dagger}_{j}\mathbb{H}_{\phi_h}\op l_{j}
+h.c.\,.
\end{split}
\end{equation}

\section{Quantum noise character in Balanced Homodyne readout}
\label{sec:qnbhd}
In this section, we focus on the effect the HOMs have on the quantum noise of an interferometer with BHD readout. We denote the input light fields at the dark port (DP) and bright port (BP) of the interferometer $\op i$ and $\op p$, respectively. Then $\op o$ and $\op q$ stand for the respective output fields. Those will contribute to the signal and LO light fields. 
Then we can introduce the I/O relations by defining the interferometer transfer matrix (TM):
\begin{widetext}
\begin{equation}\label{eq:I/O-rels_general}
\begin{split}
\underbrace{\left (
\begin{array}{c}
\op {o}_{0} \\
 \op {o}_{1} \\
 \vdots \\
\end{array}
\right)}_{\vec{\op O}}
=\underbrace{\left(
\begin{array}{ccc}
\mathbb{A}_{00} & \mathbb{A}_{01} &\cdots  \\
\mathbb{A}_{10} & \mathbb{A}_{11} &\cdots  \\
\vdots & \vdots &\ddots   \\
\end{array}
\right)}_{\mathbb{A}:\,\rm DP\to DP\
 TM} 
\left(
\begin{array}{c}
\op{i}_{0} \\
 \op{i}_{1} \\
 \vdots  \\
\end{array}
\right)
+\underbrace{\left(
\begin{array}{ccc}
\mathbb{B}_{00} & \mathbb{B}_{01} &\cdots  \\
\mathbb{B}_{10} & \mathbb{B}_{11} &\cdots  \\
\vdots & \vdots &\ddots   \\
\end{array}
\right)}_{\mathbb{B}:\,\rm BP\to DP\
 TM} 
\left (
\begin{array}{c}
\op{p}_{0} \\
 \op{p}_{1} \\
 \vdots  \\
\end{array}
\right)
+
\underbrace{\left(
\begin{array}{cccc}
\mathbf{E}_{01} & \mathbf{E}_{02} &\cdots &\mathbf{E}_{0N} \\
\mathbf{E}_{11} & \mathbf{E}_{12} &\cdots & \mathbf{E}_{1N} \\
\vdots & \vdots & &\vdots   \\
\end{array}
\right)}_{\mathbb{R}_o} 
\left (
\begin{array}{c}
{x}_{1} \\
 {x}_{2} \\
 \vdots  \\
  x_N
\end{array}
\right)
\end{split}
\end{equation}

\begin{equation}\label{eq:BPIOrels}
\begin{split}
\underbrace{\left (
\begin{array}{c}
\op {q}_{0} \\
 \op {q}_{1} \\
 \vdots \\
\end{array}
\right)}_{\vec{\op Q}}
=\underbrace{\left(
\begin{array}{ccc}
\mathbb{C}_{00} & \mathbb{C}_{01} &\cdots  \\
\mathbb{C}_{10} & \mathbb{C}_{11} &\cdots  \\
\vdots & \vdots &\ddots   \\
\end{array}
\right)}_{\mathbb{C}:\,\rm DP\to BP\
 TM}
 \left (
\begin{array}{c}
\op{i}_{0} \\
 \op{i}_{1} \\
 \vdots  \\
\end{array}
\right)
+\underbrace{\left(
\begin{array}{ccc}
\mathbb{D}_{00} & \mathbb{D}_{01} &\cdots  \\
\mathbb{D}_{10} & \mathbb{D}_{11} &\cdots  \\
\vdots & \vdots &\ddots   \\
\end{array}
\right)}_{\mathbb{D}:\,\rm BP\to BP\
 TM}
 \left (
\begin{array}{c}
\op{p}_{0} \\
 \op{p}_{1} \\
 \vdots  \\
\end{array}
\right)
+
\underbrace{\left(
\begin{array}{cccc}
\mathbf{F}_{01} & \mathbf{F}_{02} &\cdots &\mathbf{F}_{0N} \\
\mathbf{F}_{10} & \mathbf{F}_{11} &\cdots & \mathbf{F}_{1N} \\
\vdots & \vdots & &\vdots   \\
\end{array}
\right)}_{\mathbb{R}_q}
 \left (
\begin{array}{c}
{x}_{0} \\
 {x}_{1} \\
 \vdots  \\
  x_N
\end{array}
\right)
\end{split}
\end{equation}
\end{widetext}
Any single transfer matrix, e.g.\ $\mathbb{A}_{kj}$, is a $2 \times 2$ matrix that stands for the transformation from the \TEM{j} input field to the \TEM{k} output field. $\mathbf{E}_{kj}$ is a two dimensional vector of the opto-mechanical response functions of the $k$-th output mode at the DP to the displacement, $x_j(\Omega)$, of the $j$-th mechanical degree of freedom of the interferometer with $N$ being the total number of mechanical degrees of freedom. The BP response function $\mathbf{F}_{kj}$ is defined in the same way. We use notations $\mathbb{R}_o$ and $\mathbb{R}_q$ for the whole response matrices for DP and BP, respectively. 
\begin{figure*}
\centering
  \includegraphics[width=1.7\columnwidth]{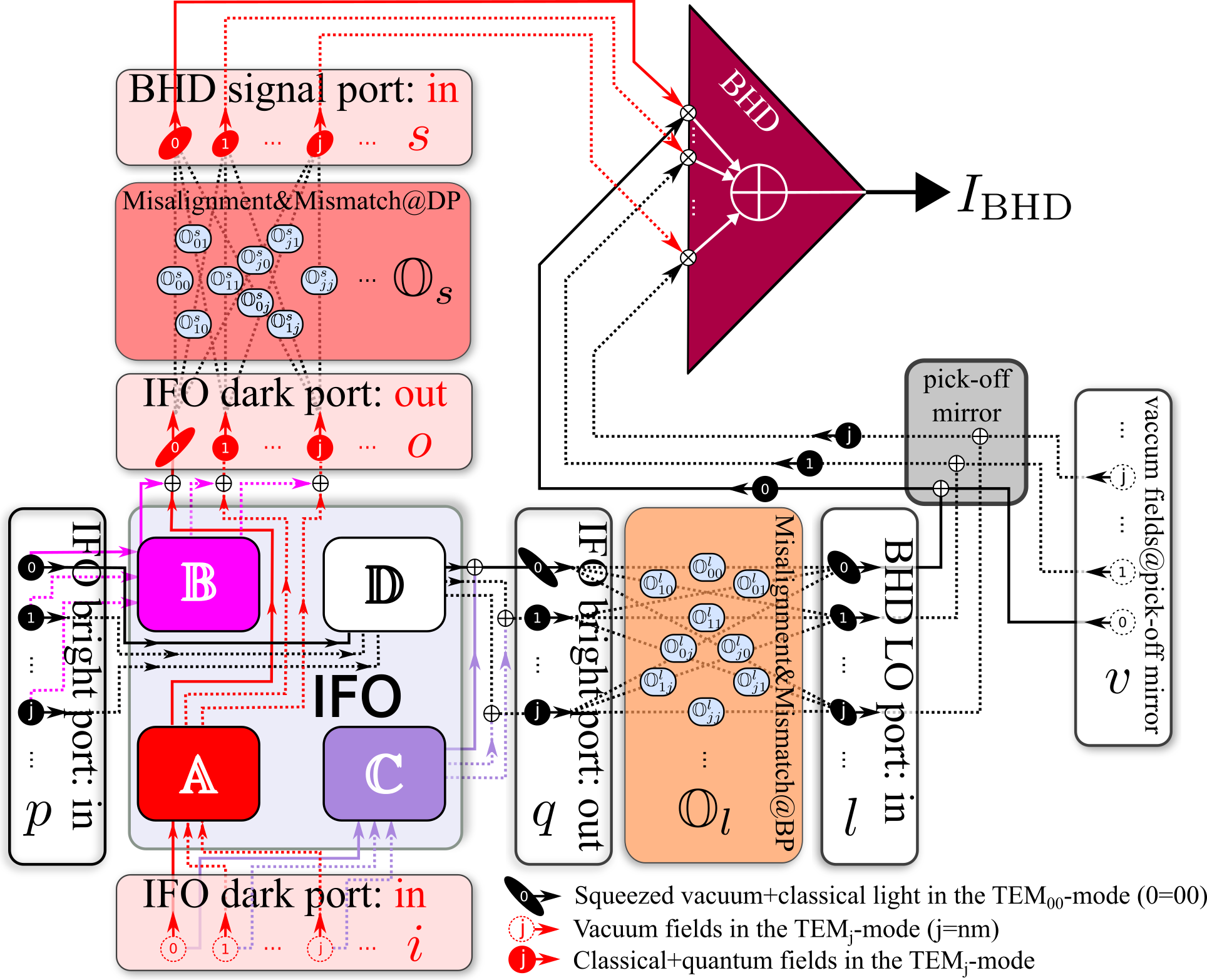}
\caption{Schematic of the HOM fields transformation in the interferometer with BHD readout. Multiple modes field $\op p$ and $\op i$ can enter to the interferometer from BP and DP, and only the interferometer mode field will suffer the ponderomotive squeezing effect, which can be explained by  the four transfer matrices  $\mathbb{A},\mathbb{B},\mathbb{C},\mathbb{D}$. The output fields from BP and DP are $\op q$ and $\op o$.  We represent the misalignment and mismatch in both paths, signal and LO, by a 
separate block, i.e. $\mathbb{O}_s$ and $\mathbb{O}_l$. The necessity for a pick off mirror in order to create the LO beam, causes additional vacuum noise $\op v$ to couple in to the BHD readout.} 
\label{fig:SSMBHD}
\end{figure*}

Since gravitational waves couple to the differential degree of freedom the arm cavities, it is sufficient for us to consider only the longitudinal motion of the two end test masses, i.e.\ $x_1$ and $x_2$, defining their common mode $x_+$  and differential mode $x_-$ via
\begin{equation}
x_1=\frac{x_{+}+x_{-}}{2}, \quad x_2=\frac{x_{+}-x_{-}}{2}.
\end{equation}
Then the response functions $\mathbf{E}_{01}$ and $\mathbf{E}_{02}$ for the fundamental light mode we measure can be written in terms of the latter ones, $\mathbf{R}_{+}$ and $\mathbf{R}_{-}$, as
\begin{equation}
\mathbf{E}_{01}=\mathbf{R}_{+}+\mathbf{R}_{-},\quad\mathbf{E}_{02}=\mathbf{R}_+-\mathbf{R}_{-},
\end{equation}

The output fields $\op o$ and $\op q$ are sent towards the BHD through a train of steering optics. The LO beam can be derived from various sources. For example, in the particular case of the Glasgow SSM (that will be discussed in detail later in this article),  the reflection from the interferometer  is used to provide the LO for the BHD, i.e. the BP as shown in Fig.~\ref{fig:BHD} and \ref{fig:SSMBHD}. Note this scenario is  more general than the simpler case of getting the LO beam by picking off some light from the pumping laser directly (by turning the beamsplitter after the laser by 90 degrees in Figure 1), for mathematically this amounts to setting to zero all $\mathbb{C}_{ij}$ and $\mathbf{F}_{ij}$ in Eq.~\eqref{eq:BPIOrels}, and also setting $\mathbb{D}_{ij} = \mathbb{I}_2\,\delta_{ij}$ with $\mathbb{I}_2$ being a $2\times2$ identity matrix and $\delta_{ij}$ the Kronecker delta. 

Due to imperfect optics, and alignment fluctuations originating from residual pendulum motion, $\op o$ and $\op q$ will suffer from misalignment and mismatch with respect to the interferometer modes. A redistribution of different modes will ensue and the new modes of the LO, $\op l$, and signal beam, $\op s$, will be a mixture of the original modes $\op o$ and $\op q$. Mismatch and misalignment can be described by scattering matrices $\mathbb{O}_l$ and $\mathbb{O}_s$ for LO beam and signal beam, respectively, defined as:
\begin{equation}\label{eq:slfields}
\begin{split}
 \left (
\begin{array}{c}
\op {s}_{0} \\
 \op {s}_{1} \\
 \vdots \\
\end{array}
\right)
= \underbrace{\left(
\begin{array}{ccc}
\mathbb{O}^{s}_{00} & \mathbb{O}^{s}_{01} &\cdots  \\
\mathbb{O}^{s}_{10} & \mathbb{O}^{s}_{11} &\cdots  \\
\vdots & \vdots &\ddots   \\
\end{array}
\right)}_{\mathbb{O}_s}
 \left (
\begin{array}{c}
\op{o}_{0} \\
 \op{o}_{1} \\
 \vdots  \\
\end{array}
\right),
\\
\left (
\begin{array}{c}
\op {l}_{0} \\
 \op {l}_{1} \\
 \vdots \\
\end{array}
\right)
= \underbrace{\left(
\begin{array}{ccc}
\mathbb{H}_{\phi_h}\mathbb{O}^{l}_{00} & \mathbb{H}_{\phi_h}\mathbb{O}^{l}_{01} &\cdots  \\
\mathbb{H}_{\phi_h}\mathbb{O}^{l}_{10} & \mathbb{H}_{\phi_h}\mathbb{O}^{l}_{11} &\cdots  \\
\vdots & \vdots &\ddots   \\
\end{array}
\right)}_{\mathbb{O}_l}
 \left (
\begin{array}{c}
\op{q}_{0} \\
 \op{q}_{1} \\
 \vdots  \\
\end{array}
\right),
\end{split}
\end{equation}
where (as shown in Fig.~\ref{fig:SSMBHD}) the matrix component $\mathbb{O}^{s}_{kj}$ ($\mathbb{O}^{l}_{kj}$) describes how the $j$-th mode of the $\op o$ field ($\op q$ field) contributes to the $k$-th mode of the $\op s$ field ($\op l$ field). Each $\mathbb{O}^{s}_{kj}$ ($\mathbb{O}^{l}_{kj}$) is a $2 \times 2$ matrix.  $\mathbb{O}^{s}$ ($\mathbb{O}^{l}$) are not arbitrary, rather they need to satisfy the unitarity relation ${\mathbb{O}_s^{\dagger}}\mathbb{O}_s ={\mathbb{O}^{\dagger}_l}\mathbb{O}_l=\mathbb{I}$, where $\mathbb{I}$ is the identity matrix, as a consequence of the law of energy conservation. 

As the LO field mixes in a vacuum field $\op v$ coming from the open port of the pick-off mirror (see Fig.~\ref{fig:SSMBHD}), the actual LO field at the BHD reads 
\begin{equation}
\op l'=\sqrt{R_{p}}\op l+\sqrt{T_{p}} \op v \,,
\end{equation}
where $R_{p}$ and $T_{p}$ are the power reflectivity and transmissivity of the pick-off mirror, respectively. 
Then according to Eq.~\eqref{eq:Ifluctuation}, we can write out the BHD readout photocurrent in terms of quantum noise and differential mode motion as
\begin{widetext}
\begin{equation}\label{eq:GIbhd}
\begin{split}
I_{\rm BHD}\propto \sqrt{R_{p}} \op L^{\dagger}  \bigl[\mathbb{O}_s ( \mathbb{A} \op i+\mathbb{B}  \op p  ) \bigl]  
+ \op S^{\dagger}\bigr[\sqrt{R_p} \mathbb{O}_l ( \mathbb{C} \op i+\mathbb{D}  \op p  )+\sqrt{T_{p}}\op v\bigr] 
+ \sqrt{R_{p}}\op L^{\dagger}   \left(
\begin{array}{c}
\mathbb{O}^s_{00} \\
\mathbb{O}^s_{10} 
\\ 
\vdots
\end{array}
\right) \ \mathbf{R}_{-} \ x_{-}+h.c.\,,
\end{split}
\end{equation}
\end{widetext} 
in which we neglect the term $\propto \op l^{\dagger}\op s$, the second order term in the noise fluctuations. Finally, using the formalism of Eq.~(12) in \cite{Danilishin2015}, one can write down the quantum noise power spectral density as
\begin{equation}\label{eq:Spectraldensity}
\begin{split}
S\propto \Bigl[\op L^{\dagger}  \mathbb{O}_s  (\mathbb{A}\,  \mathbb{S}_i  \, \mathbb{A}^{\dagger}+\mathbb{B}\,   \mathbb{S}_p\,   \mathbb{B}^{\dagger})   \mathbb{O}_s^{\dagger}  \op L 
\\
+\op {S}^{\dagger} \mathbb{O}_l (\mathbb{C}\,   \mathbb{S}_i\,   \mathbb{C}^{\dagger}+ \mathbb{D}\,   \mathbb{S}_p\,   \mathbb{D}^{\dagger}) \mathbb{O}_l^{\dagger}  \op S
\\
+\op L^{\dagger} \mathbb{O}_s (\mathbb{A}\,   \mathbb{S}_i\,   \mathbb{C}^{\dagger}+\mathbb{B}\,   \mathbb{S}_p\,   \mathbb{D}^{\dagger})  \mathbb{O}_l^{\dagger}  \op S 
\\
+\op S^{\dagger} \mathbb{O}_l (\mathbb{C}\,   \mathbb{S}_i \, \mathbb{A}^{\dagger}+\mathbb{D}\,   \mathbb{S}_p\,   \mathbb{B}^{\dagger})  \mathbb{O}_s^{\dagger}  \op L  
\\
+\frac{T_{p}}{R_{p}}\op S^{\dagger} \op S \Bigr]
\\
/\left| \op {L}^{\dagger}\left(
\begin{array}{c}
\mathbb{O}^s_{00} \\
\mathbb{O}^s_{10} 
\\ 
\vdots
\end{array}
\right)  \mathbf{R}_{-} \right|^2
\end{split} \,,
\end{equation}
where $\mathbb{S}_i$ and $\mathbb{S}_p$ are the power spectral density matrices of $\op i$ and $\op p$ input fields \footnote{Note that the dimensions of $\mathbb{S}_i$ and $\mathbb{S}_p$ correspond to the number of optical modes taken into consideration. In general, they have the same dimensions as TMs.}. For each optical mode, the components of $\mathbb{S}_i$ and $\mathbb{S}_p$ are defined as
\begin{equation}
\begin{split}
\pi \mathbb{S}_{jj'}^{i} \delta_{jj'}\delta(\Omega-\Omega')\equiv 
\langle {\hat{i}_{j}}(\Omega){\hat{i}_{j'}^{\dagger}}({\Omega'})+\hat{i}^{*}_j(\Omega')\hat{i}^{\rm T}_{j'}(\Omega)
\rangle
\\
\pi \mathbb{S}_{jj'}^{p} \delta_{jj'}\delta(\Omega-\Omega')\equiv 
\langle {\hat{p}_{j}}(\Omega){\hat{p}_{j'}^{\dagger}}({\Omega'})+\hat{p}^{*}_j(\Omega')\hat{p}^{\rm T}_{j'}(\Omega)
\rangle
\end{split}\,,
\end{equation}
where we define the hermitian conjugate of the two dimensional vector of light quadratures of the $j$-th mode as ${\op i_j}^{\dagger}=(\op i_{c,j}^\dag,\op i_{s,j}^{\dagger})$ and the complex conjugate of the same vector as ${\op i_j}^{*}=(\op i_{c,j}^{\dag},\op i_{s,j}^{\dag})^{\rm T}$. $\mathbb{S}_{jj'}^{i,p}$ are $2 \times 2$ matrices of power spectral densities of input fields in the $j$-th mode when $j=j'$ and cross spectral densities between the $j$-th and $j'$-th modes of the corresponding input fields, if there are any.  

\section{Mathematical treatment of mismatch and misalignment of multi-mode Hermite-Gaussian beams in a linear optical setup }
\label{sec:couplingc}

In this section, following the formalism of \cite{BayerHelms84}\cite{Bond2009}, we calculate scattering matrices that describe transformation of the multi-mode Gaussian beam as it undergoes misalignment in the imperfect optical steering train from an input or output port of the interferometer to the corresponding input port of the balanced homodyne detector.       

Firstly, we define the Cartesian coordinate system $(x,y,z)$ for the mode at the output port of the interferometer. We assume the beam propagates along the $z$-axis with $z_0$ being the position of the beam waist and $z=0$ is the location of the observation plane. $x$, $y$ are the transverse spatial coordinates. Then, the spatial profile of the Hermite-Gaussian beam is given by
\begin{widetext}
\begin{equation}
u_{mn}(x,y,z)=(1+{\zeta} ^2)^{\frac{1}{2}}N_{mn}
    H_m\Bigl[\bigl(\frac{2}{1+\zeta ^2}\bigr)^{\frac{1}{2}}\frac{x}{w_0}\Bigr]
    H_n\Bigl[\bigl(\frac{2}{1+\zeta ^2}\bigr)^{\frac{1}{2}}\frac{y}{w_0}\Bigr]
    e^{-i kz+i(m+n+1)\arctan (\zeta) -\frac{x^2+y^2}{w_0^2(1-i\zeta)}}\,,
\end{equation}
\end{widetext}
where $\zeta=\frac{z-z_0}{z_R}$ is a normalized $z$-coordinate and $z_R$ is the Rayleigh range of the beam.  We define the angular aperture of the beam as $\gamma=w_0/z_R$. The normalization factor $N_{mn}$ is given by $N_{mn}=\pi w_0^2\,2^{m+n-1}m!n!$. The individual modes satisfy the orthogonality condition
\begin{equation}\label{eq:ort}
 \iint_{-\infty}^{+\infty}dr\, {u_{mn}(x,y,z)}{u_{kl}^{*}(x,y,z)}=\delta_{mk}\delta_{nl} \,. \\
\end{equation}

\begin{figure}[bt]
\centering
  \includegraphics[width=1\columnwidth]{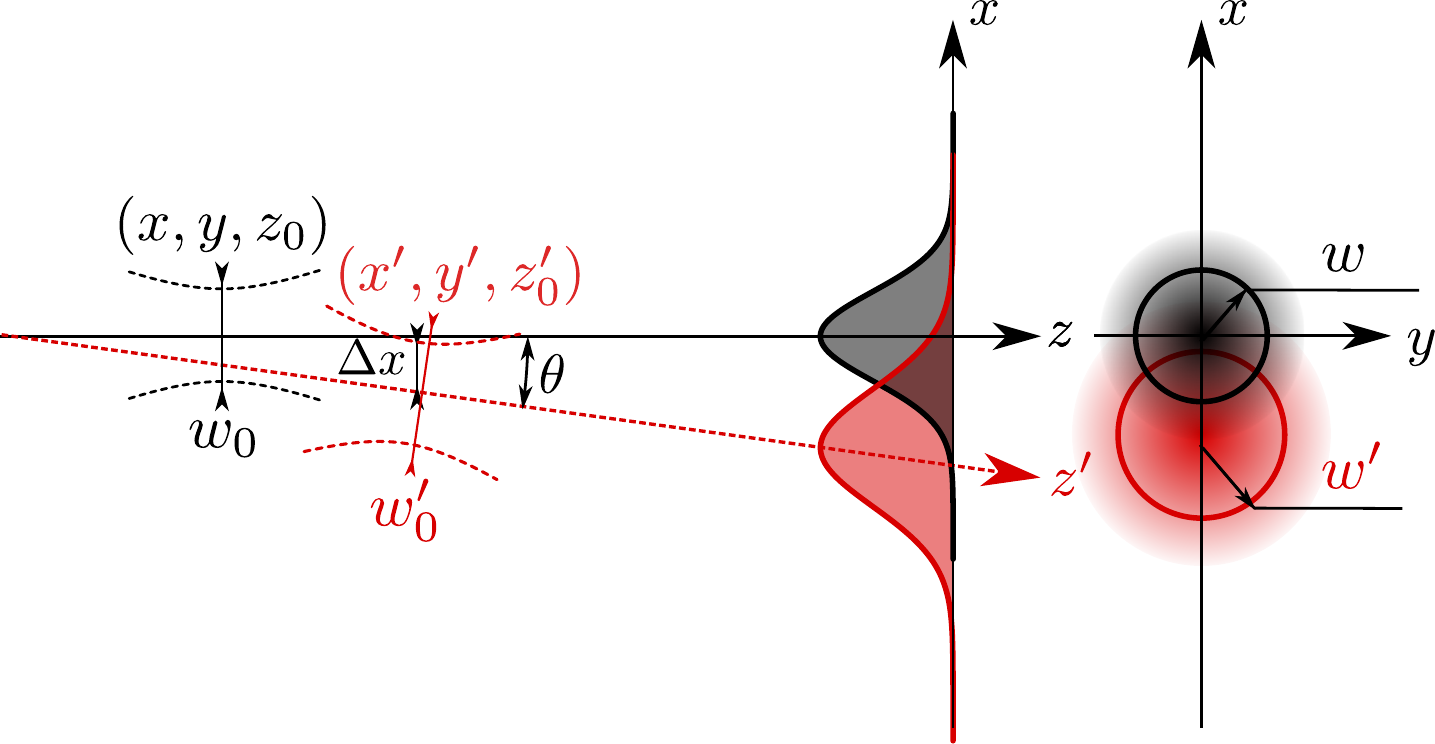}
\caption{Schematic of the general mismatch and misalignment transformation of the Gaussian beam. The waist sizes  
of the initial beam and the transformed beam are given by $w_{0}$, $w_{0}'$, respectively. $z_0$ and $z'_0$ stand for the coordinates of the waist position of the two beams in the corresponding coordinate systems. The observation plane is located at $z=0$ and $z'=0$. The misalignment can be described by the angular misalignment $\theta$, as well as by the displacements $\Delta x$ and $\Delta y$.} 
\label{fig:misalignment_BHD}
\end{figure}

We then introduce a misalignment of the beam by an angle $\theta$ around the $-y$-axis at the beam waist location, followed by transverse displacements $\Delta x$ and $\Delta y$. These transformations yield the new misaligned beam coordinate system $(x',y',z')$ (see Fig.~\ref{fig:misalignment_BHD}). In addition, we allow for a mismatch of the beam parameters, which can be described by the two coefficients
\begin{equation}
\begin{aligned}
K_0 &= \frac{z_0-z'_0}{z_R}\,, \\ 
K_R &= \frac{z'_R-z_R}{z_R}=\frac{w_0'^2-w_0^2}{w_0^2}\,. \\ 
\end{aligned}
\end{equation}
Therefore, misalignment of the two beams is parametrised by $\Delta x$, $\Delta y$ and $\theta$, while the mismatch in beam size and wavefront curvature is parametrised by $K_0$ and $K_R$. The transformation between $(x,y,z)$ and $(x',y',z')$ can then be written as
\begin{equation}\label{eq:tfor}
\begin{aligned}
\frac{x'}{w'_0} &= \frac{x + \Delta x +z\sin(\theta)}{(1+K_R)^{\frac{1}{2}}w_0}\,,\\
\frac{y'}{w'_0} &= \frac{y + \Delta y}{(1+K_R)^{\frac{1}{2}}w_0}\,,\\
\zeta'&=\frac{\zeta+K_0}{1+K_R}\,.
\end{aligned}
\end{equation}

As the spatial modes of the initial beam, $u_{mn}(x,y,z)$, comprise a full orthonormal set, any mode $u_{m'n'}$ of the misaligned beam can be expressed in terms of the former,
\begin{equation}
u_{m'n'}(x',y',z')=\sum _{m=0}^{\infty} \sum _{n=0}^{\infty}c^{m'n'}_{mn}u_{mn}(x,y,z)\,.
\end{equation}
The coupling coefficients $c^{m'n'}_{mn}$ are obtained from Eqs.~\eqref{eq:ort} and \eqref{eq:tfor}, resulting in
\begin{multline}
c^{m'n'}_{mn}\\=e^{-ik(z'_0-z_0)}e^{i2kz\sin^2(\frac{\theta}{2})} 
\iint_{-\infty}^{+\infty}dr\,
u_{m'n'} u_{mn}^{*} e^{ikx\sin(\theta)}.
\end{multline}
Since Hermite-Gaussian modes are factorisable in $x$ and $y$, the same applies to the coupling coefficients, \textit{i.e.} $c^{m'n'}_{mn} =c_{m}^{m'} c_{n}^{n'}$. According to \cite{BayerHelms84}, the factorised coupling coefficient reads:
\begin{widetext}
\begin{subequations}
\label{eq:coeffs}
\begin{align}
\label{eq:coeff}
c^{m'}_{m} &= (-1)^m E^{(x)} \bigl(
        m'!m! (1+K_R)^{m'+\frac{1}{2}} (1+K^*)^{-(m+m'+1)}
    \bigr)^{\frac{1}{2}}
    [S_g-S_u]e^{\frac{-ik(z'_0-z_0)}{2}}\,, \\
\label{eq:Sg}
S_g &=
    \sum_{\mu'=0}^{[m'/2]}
    \sum_{\mu =0}^{[m/2]}
    \frac{(-1)^{\mu'}X^{m'-2\mu'}{X}'^{m-2\mu}}{(m'-2\mu')!(m-2\mu)!}
    \sum_{\sigma=0}^{\min(\mu, \mu)}
    \frac{(-1)^{\sigma}F^{\mu'-\sigma}{F}'^{\mu-\sigma}}{(2\sigma)!(\mu'-\sigma)!(\mu-\sigma)!} \,, \\
\label{eq:Su}
S_u &=
    \sum_{\mu'=0}^{[(m'-1)/2]}
    \sum_{\mu=0}^{[(m-1)/2]}
    \frac{(-1)^{\mu'}X^{m'-2\mu'-1}{X}'^{m-2\mu-1}}{(m'-2\mu'-1)!(m-2\mu-1)!}
    \sum _{\sigma=0}^{\min(\mu',\mu)}
    \frac{ (-1)^\sigma F^{\mu'-\sigma}{F}'^{\mu-\sigma}}{(2\sigma+1)!(\mu'-\sigma)!(\mu-\sigma)!}.
\end{align}
\end{subequations}
\end{widetext}

The symbol $[m/2]$ stands for the integer part of $\frac{m}{2}$. $S_u=0$ for $m=0$ or $m'=0$. The notations in Eqs.~\eqref{eq:coeffs} are given in  Table~\ref{ta:par}. For the $y$ axis, $m$, $m'$ have to be replaced by $n$, $n'$ and $X$, $X'$ by $Y$. 

\begin{table}[!htbp]
\caption{Notations used in Eqs.~\eqref{eq:coeffs}.} 
\label{ta:par}
\begin{ruledtabular}
\begin{tabular}{cc}
    $K$ & $\frac{K_{R}+iK_{0}}{2}$ \\
    $X$ & $(1+K^*)^{-\frac{1}{2}}(\frac{\Delta x}{w_0}-(\frac{(-z_0)}{z_R}-i)\frac{\theta}{\gamma}) $\\
    $X'$ & $(1+K^*)^{-\frac{1}{2}}(\frac{\Delta x}{w_0}-(\frac{(-z_0')}{z_R}+i(1+2K^*))\frac{\theta}{\gamma})$\\
    $Y$ &$(1+K^*)^{-\frac{1}{2}}\frac{\Delta y}{w_0}$\\
    $F$ &$\frac{K}{2(1+K_{R})}$\\
    $F'$ &$\frac{K^*}{2}$\\
    $E^{(x)}$ &$e^{-\frac{X'X}{2}-i\frac{\Delta x}{w_0}\frac{\theta}{\gamma_0}}$\\
    $E^{(y)}$ & $e^{-\frac{y^2}{2}}$\\
\end{tabular}
\end{ruledtabular}
\end{table}
As misalignment angles and shifts are usually small compared to the wave front curvature scale, hereafter we neglect the effect of wave front tilting.       

The above calculated coefficients can be translated into the components of the scattering matrices $\mathbb{O}^{s}_{jj'}$ and $\mathbb{O}^{l}_{jj'}$, which describe the misalignment effects in the signal and LO path, in two-photon formalism, for the individual optical modes at the corresponding input ports of the BHD:

\begin{equation}\label{eq:C-M}
\mathbb{O}^{s}
_{kj}=\left|c^k_j\right| \mathbb{H}_{\phi_{kj}}\,,\quad\mathbb{O}^{l}
_{kj}=\left|d^k_j\right| \mathbb{H}_{\psi_{kj}} 
\end{equation}
where $c^k_j \to c^{m'n'}_{mn} = c^{m'}_{m}c^{n'}_{n} $ and $\phi_{kj} \equiv \arg(c^k_j)$, and similarly for $d^k_j$ and $\psi_{kj}$.  

The many elements in the optical paths that connect the output ports of the interferometer to the corresponding input port of the BHD each apply their own misalignment and mismatch transformations. Here we reduce this complexity to a single effective beam rotation ($\theta$), lateral ($\Delta x$ and $\Delta y$) shifts of the beam and modified beam parameters ($z_0'$, $z_R'$) as they are measured at the detection point, i.e.\ at the input of the BHD. It can be easily shown that this does not undermine the generality of our treatment, and the transform that any linear optical system does to the HG optical beam can be represented in that way \cite{BayerHelms84,siegman1986}. 

\begin{figure}[bt]
\centering
  \includegraphics[width=1\columnwidth]{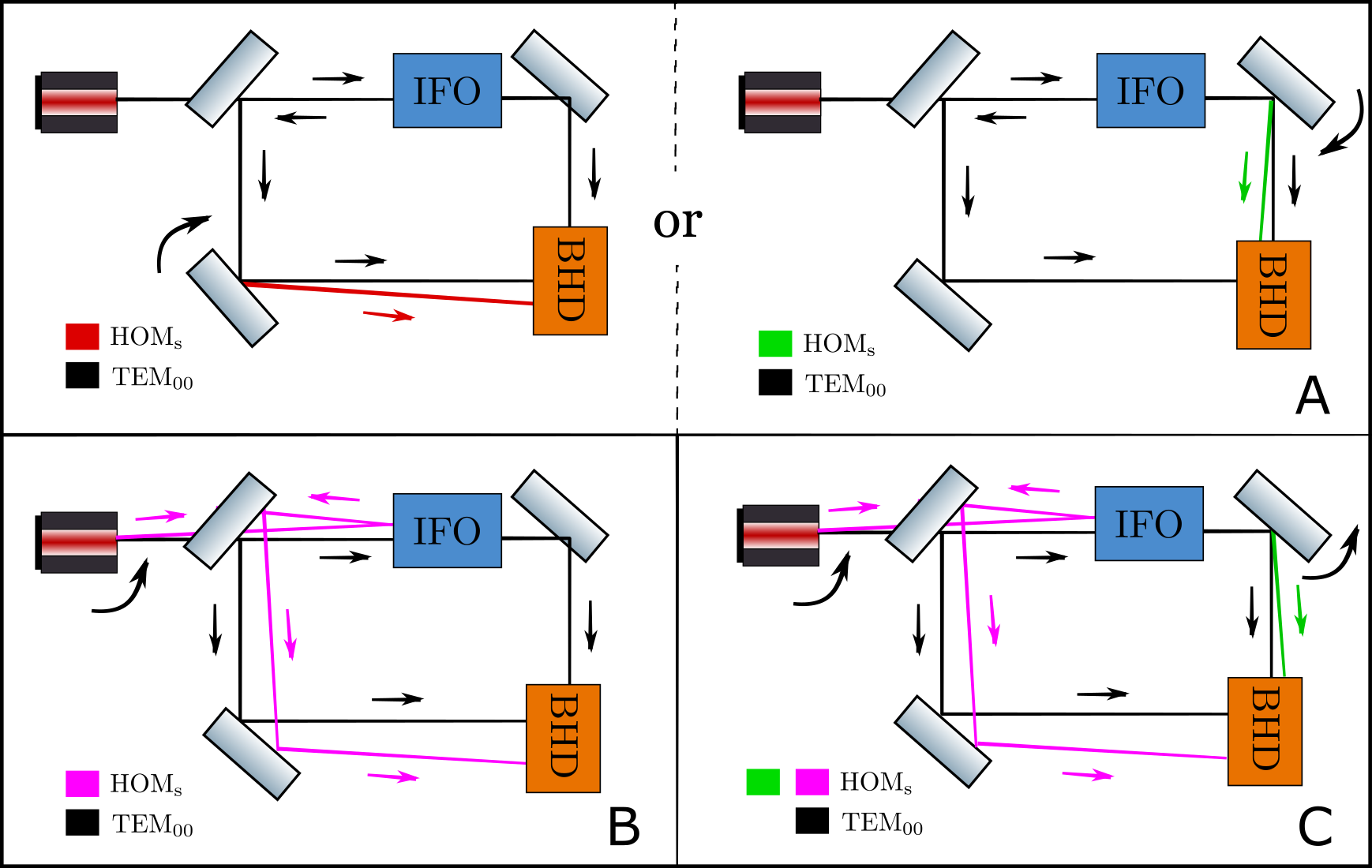}
\caption{Schematic of the output ports and input ports misalignment. The black line indicates the fundamental mode defined by the arm cavities of the interferometer. The coloured lines show the HOMs components caused by different misalignment conditions. A)  \emph{output misalignment}, \textit{i.e.} misaligned LO path, or  misaligned signal path, respectively; B)  \emph{input misalignment}, \textit{i.e.} misaligned input laser beam, which will contribute HOMs to LO beam and reduce power inside the main interferometer; C) combination of input port misalignment and output signal port misalignment.} 
\label{fig:misalignment_condition}
\end{figure}

\section{Influence of higher-order modes on the quantum noise in a Michelson interferometer with Balanced Homodyne Detection}
\label{sec:MichelsonBHD}

In this section, we provide the application of above framework on the conventional Fabry-P\'erot--Michelson interferometer. The interferometer transfer matrices, $\mathbb{A}$, $\mathbb{B}$, $\mathbb{C}$, $\mathbb{D}$, defined in Eqs.~\eqref{eq:I/O-rels_general} can be written for our particular case as~\cite{kimble2001}.
\begin{subequations}
\begin{equation}
\mathbb{A}_{00}=
	e^{2i\beta_{\rm arm}}\begin{bmatrix}
	1 & 0 \\
	-\mathcal{K}_{\rm MI} & 1
	\end{bmatrix},\,
\mathbb{B}_{00}=
	\begin{bmatrix}
	0 & 0\\
	0 & 0
	\end{bmatrix},
\end{equation}

\begin{equation}
\mathbb{C}_{00}=
	\begin{bmatrix}
	0 & 0\\
	0 & 0
	\end{bmatrix},\,
\mathbb{D}_{00}= e^{2i\beta_{\rm arm}}
	\begin{bmatrix}
	1 & 0\\
	-\mathcal{K}_{\rm MI} & 1
	\end{bmatrix}
\end{equation}
\end{subequations}
for the fundamental mode of the interferometer. $\mathcal{K}_{\rm MI}$ is the opto-mechanical coupling factor of a Fabry-P\'erot--Michelson interferometer defined as:
\begin{equation}
\mathcal{K}_{\rm MI}=\frac{2\Theta \gamma_{arm}}{\Omega^2(\gamma_{arm}^2+\Omega^2)^2}~,
\end{equation}
where $\gamma_{arm}=\frac{cT_{ITM}}{4L}$ is the half-bandwidth of the arm cavities of length $L$  and with input mirror power transmittance $T_{ITM}$, and $\Theta = \frac{4\omega P_{arm}}{McL}$ is the normalised circulating power in both arms. 

For the HOMs, i.e.\ for $j,k>0$, we assume the high-finesse arm-cavity interferometer to be a highly selective mode filter that does not let HOMs in, rather reflecting them off without  any dispersion (frequency dependent phase shift). Therefore the corresponding transfer matrices take a particularly simple form:  

\begin{align}
\mathbb{A}_{kj} = \mathbb{D}_{kj} &=\delta_{kj}
\begin{bmatrix}
1 & 0\\
0 & 1
\end{bmatrix}\,,
& \mathbb{B}_{kj} = \mathbb{C}_{kj} =
\begin{bmatrix}
0 & 0\\
0 & 0
\end{bmatrix}\,,
\end{align}
indicating that the vacuum noise in HOMs is reflected to the output port right away, without any additional phase shift. However, the fundamental mode light interacts with the interferometer and thereby it gets ponderomotively squeezed by the opto-mechanical interaction with the mechanical degrees of freedom of the interferometer. This fact is reflected in Fig.~\ref{fig:misalignment_BHD} by squeezed error ellipse of the TEM$_{00}$ mode at both IFO output ports, $\op{o}$ and $\op{q}$.

The response of the interferometer to the differential mechanical modes of the arm mirrors, that are of particular interest in the context of gravitational wave detectors, can be written as:

\begin{eqnarray} 
&\pmb{R}_{\rm -} &= e^{i\beta_{\rm arm}}\dfrac{\sqrt{2\mathcal{K}_{\rm MI}}}{x_{\rm SQL}}\begin{bmatrix}0\\1\end{bmatrix}\,,
\end{eqnarray}
where $x_{\rm SQL}$ stands for the single-sided spectral density of the standard quantum limit in terms of displacement, and $\beta_{\rm arm} = \arctan (\frac{\Omega}{\gamma_{arm}})$ is the phase shift that the light sidebands with frequency $\Omega$ acquire when propagating through and reflecting off the arm cavity~\cite{SD2012}.

We can distinguish three different cases of how misaligments can couple into the BHD readout:
\begin{enumerate}
\item \textbf{Output misalignment} occurring in one or both of the BHD paths, which  refers to "a" and "b" in Fig.~\ref{fig:misalignment_condition}.
\item \textbf{Input misalignment} to the interferometer, which will cause multiple mode fields to be injected into the interferometer as shown in Fig.~\ref{fig:SSMBHD} and referred to  as "c" in Fig.~\ref{fig:misalignment_condition}.
\item \textbf{Combination of the input and output misalignment}, which refers to the "d" in Fig.~\ref{fig:misalignment_condition}.  
\end{enumerate}

We note that the pick off mirror is set to pick up the reflection beam of the interferometer as LO beam. As the specific design for implementing the BHD readout in a full large scale GW detector \textit{i.e.} Advanced Ligo is still under discussion, in the following we use similar instrument parameters as for the  Fabry-P\'erot--Michelson considered in \cite{G2014}.  The input power is 3.4\, W, the power transmissivity of cavity input test mass is 700 ppm, the effective cavity mass is around 1g,  and the arm cavity length is around 1.4m. 

\subsection{Output Misalignment}
\begin{figure}[bt]
\centering
\includegraphics[width=1\columnwidth]{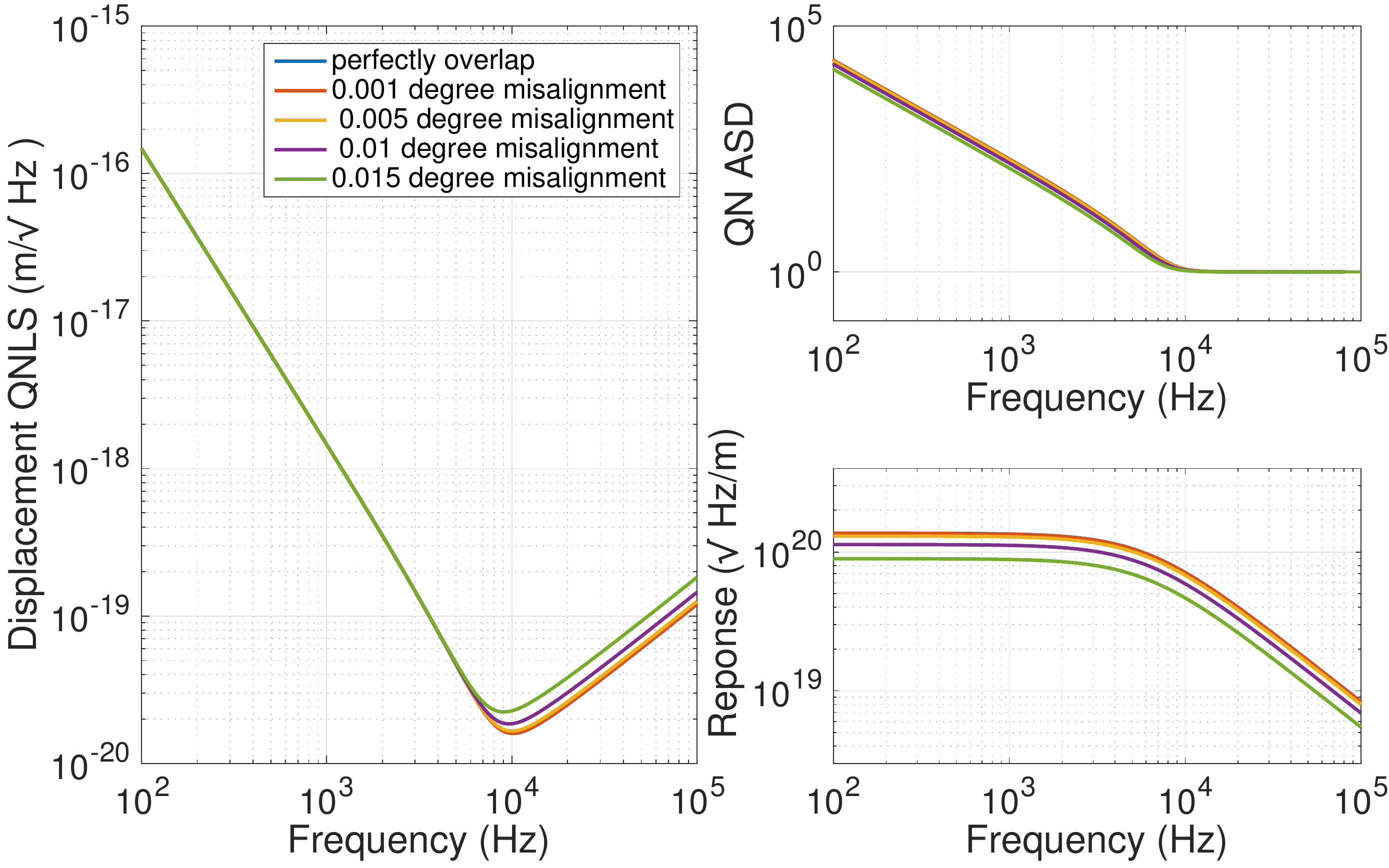}
\caption{\textit{Left panel}: Displacement quantum noise limited sensitivity (QNLS). \textit{Upper right panel}: Quantum noise (QN) amplitude spectral density. \textit{Lower right panel}: response function of the interferometer for different values of misalignment angle between the LO beam and the signal one at the BHD.  It refers to part  A) in Fig.~\ref{fig:misalignment_condition}. This gives the following values of equivalent relative lateral displacement of the two beams normalised by the beam radius on photodiode: 0.05, 0.25, 0.5, 0.7.} 
\label{fig:OutMisa} 
\end{figure}

The left hand plot of  Fig.~\ref{fig:OutMisa} shows the effect of output misalignment onto the quantum noise limited displacement sensitivity of our example Fabry-P\'erot--Michelson interferometer with BHD using a phase quadrature readout. 
The differently coloured traces indicate different magnitudes of misalignments. The right hand top plot shows the amplitude spectral density (ASD) of the quantum noise, while the lower plot on the right hand side shows the response of the differential arm length degree of the interferometer. 

For output misalignment  we  obtain that at the frequencies below 5\,kHz, where radiation pressure noise dominates in the interferometer, there is no visible influence on the quantum noise limited sensitivity due to HOMs in the BHD paths. The most pronounced effect can be seen in the shot-noise-dominated frequency band, i.e. above 5\,kHz. This can be understood by the following chain of arguments. The ponderomotive squeezing, which is described by  "$\mathcal{K}_{\rm MI}$",   is responsible for the radiation pressure noise at low frequencies and affects only the TEM$_{00}$ mode. The effect of misalignment on this mode can be  described by a simple multiplication by the factors $|d^0_0|<1$ and $|c^0_0|<1$ of the fundamental mode contributions to the LO and the signal beams, including the arm mirrors displacement signal. While the contribution of the HOMs can have in general a complicated structure at the level of field operators, the fact that all fluctuating parts of the HOM fields are in the vacuum state, which is invariant to phase shifts, the resulting additional noise in the BHD photocurrent can be described by the noise operators, $\op{n}^{\rm HOM}$ that absorb all the HOM vacuum fields and enter the readout signal with effective coefficient $\sqrt{1-|d^0_0|^2}$ and $\sqrt{1-|c^0_0|^2}$, correspondingly. Assuming that there is no   significant classical field leaving the Michelson interferometer at the dark port, one can safely neglect the noise contribution of the cross term between classical component in signal beam and quantum noise in LO beam.
Then we write out the BHD photocurrent in the phase quadrature for the case of a misaligned LO beam as 

\begin{multline}\label{eq:BHDphotos1}
I_{\rm BHD}\propto |\op{L}|(|c^0_0|\bigl[e^{2i\beta_{\rm arm}}(-\mathcal{K}_{\rm MI}\hat{i}^0_c +\hat{i}^0_s)+ e^{i\beta_{\rm arm}}\dfrac{\sqrt{2\mathcal{K}_{\rm MI}}}{x_{\rm SQL}}x_-\bigr]\\
 + \sqrt{1-|c^0_0|^2}\Delta\op{n}^{\rm HOM}_s)+h.c.
\end{multline}
where $|\op{L}|$ represents the magnitudes of the LO DC components. Analogously we can describe the case of a misaligned signal beam by replacing  $c^0_0$ with $d^0_0$.

\begin{figure}
\centering
\includegraphics[width=1\columnwidth]{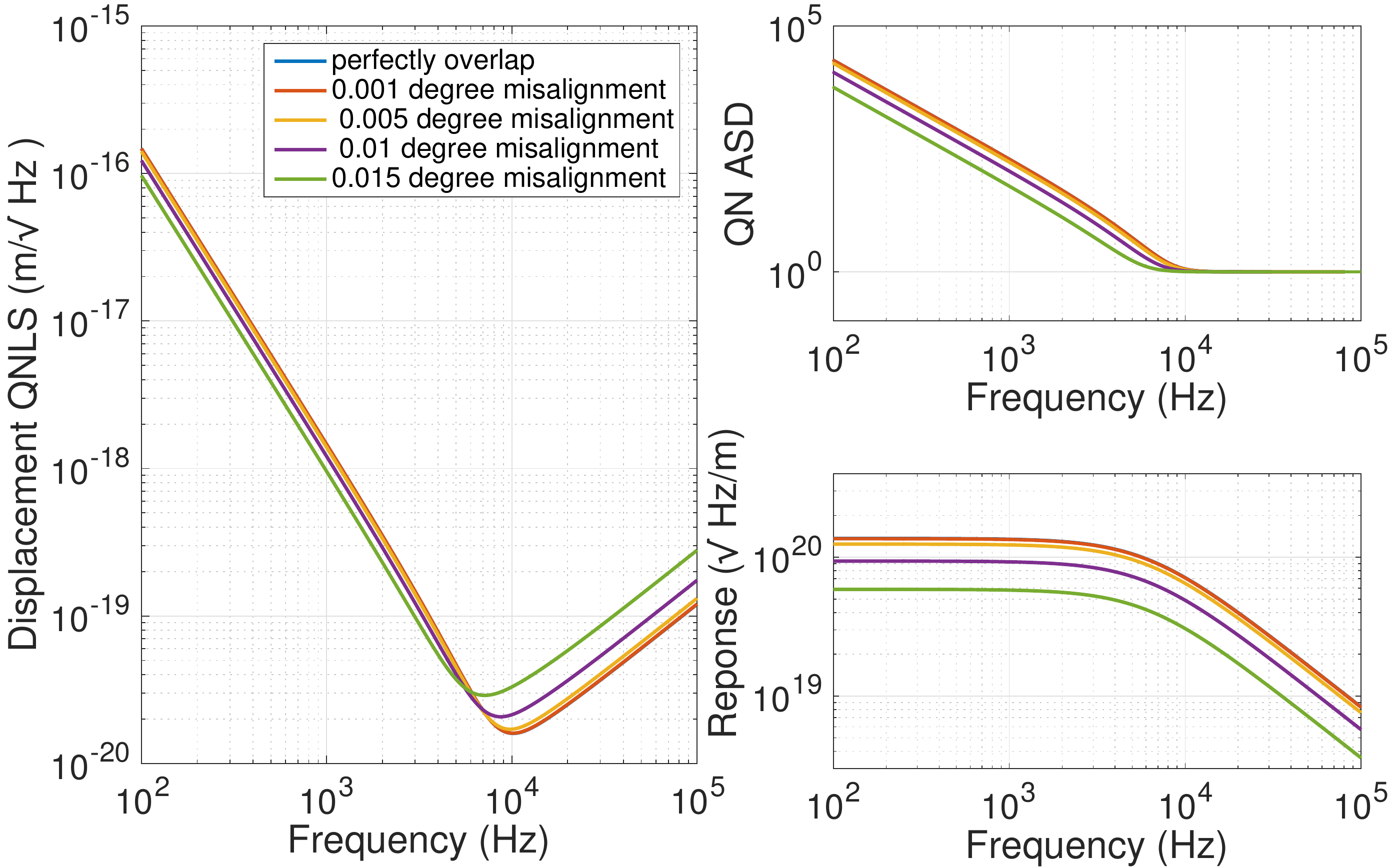}
\caption{ \textit{Left panel}: Displacement quantum noise limited sensitivity (QNLS). \textit{Upper right panel}: Quantum noise (QN) amplitude spectral density. \textit{Lower right panel}: response function of the interferometer for different values of misalignment angle between the input laser beam and the interferometer. It refers to the part B) in Fig.~\ref{fig:misalignment_condition}. } 
\label{fig:InMisa}
\end{figure}

\begin{figure}
\centering
\includegraphics[width=1\columnwidth]{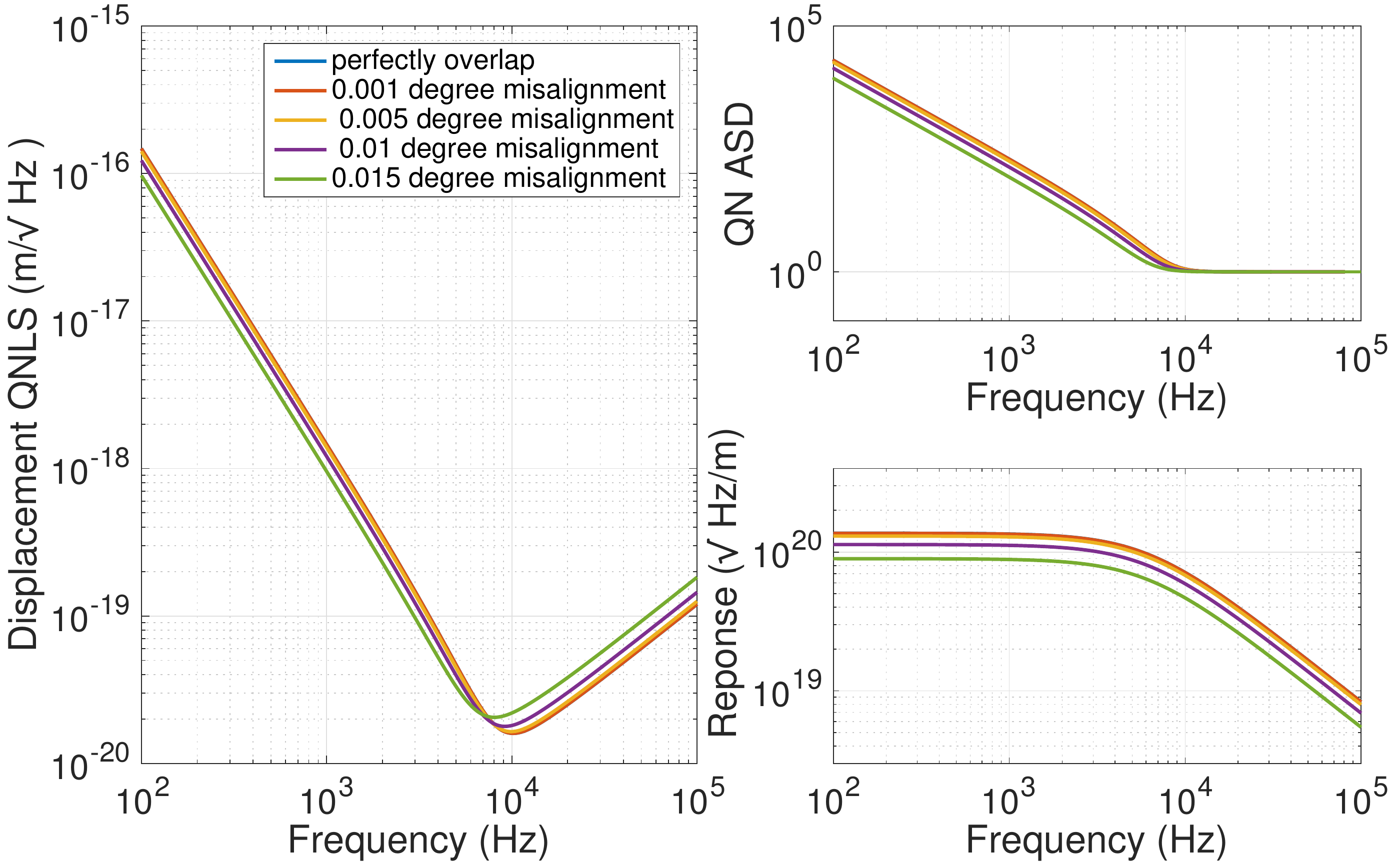}
\caption{\textit{Left panel}: Displacement quantum noise limited sensitivity (QNLS). \textit{Upper right panel}: Quantum noise (QN) amplitude spectral density. \textit{Lower right panel}: response function of the interferometer for different values of misalignment angle between the input laser beam and the interferometer, and at the same time the same amount and same direction misalignment for the signal beam is set to recover perfect overlap between signal field and LO field. It refers to the part C) of Fig.~\ref{fig:misalignment_condition}. } 
\label{fig:InOutMisa}
\end{figure}

Therefore, the effect of misalignment and HOM contamination of the readout signal is mathematically equivalent to the effect of loss at the readout photodetectors with misalignment coefficient $|c^0_0|^2$ serving as an effective quantum efficiency of the readout. Indeed, radiation pressure noise creates the real displacement of the mirrors of the interferometer indistinguishable from the signal displacement. Therefore, apparently the signal-to-noise ratio (SNR) for back-action noise is not influenced by the output beams misalignment, hence the back-action dominated part of QNLS. Shot noise, on the contrary, remains the same disregarding the level of output loss while the signal magnitude decreases proportionally. Thus SNR for shot noise goes down, worsening the QNLS, as can be seen in Fig.~\ref{fig:OutMisa} and in the following two formulae for QN and QNLS, respectively:
\begin{equation}
S^{\rm QN} \simeq |c^0_0|^2 (S^{0}_{s,\, SN}(f) + S^{0}_{s,\, BA}(f)) + (1-|c^0_0|^2)\,,
\end{equation}
where $S^{0}_{s,\, SN}(f) = 1$ and $S^{0}_{s,\, BA}(f)$ are the shot noise and back-action components of the power spectral density (PSD) of the quantum noise on phase quadrature at the dark port in TEM$_{00}$ mode, respectively, and for QNLS:
\begin{multline}
S^{\rm QN}_x \simeq \dfrac{|c^0_0|^2 (1 + S^{0}_{s,\, BA}(f))  + (1-|c^0_0|^2)}{|c^0_0|^2|\op{R}_{-s}|^2} =\\= \dfrac{ S^{0}_{s,\, BA}(f)}{|\op{R}_{-s}|^2} + \dfrac{1}{|c^0_0|^2|\op{R}_{-s}|^2}\,.
\end{multline}
where $|c^0_0|^2|\op{R}_{-s}|^2$ stands for the optomechanical response function, emphasising the signal contents reduced by $|c^0_0|^2$.

\subsection{Input Misalignment}
Fig.~\ref{fig:InMisa} shows the effect of input misalignment, which refers to part B) in Fig.~\ref{fig:misalignment_condition}. We find that in the low frequency range the sensitivity suffers more than in the case of output misalignment, while the low frequency sensitivity benefits instead. 

The effect of input misalignment is twofold: on the one hand, HOMs contaminate the local oscillator beam and lead to the decrease of the LO fundamental mode amplitude by a factor of $|e^0_0|<1$, which amounts to the same effect as described above for output misalignment. On the other hand, this also reduces the amount of classical light circulating in the fundamental mode of the interferometer by $|e^0_0|^2$, thereby reducing the back-action noise, represented by the optomechanical coupling factor $\mathcal{K}_{\rm MI}$ in the I/O-relations \footnote{Depending on the cause of input misalignment, \textit{e.g.} limited gain in auto alignment system or unwanted offset present in auto alignment system, it could be detuned by monitoring optical power inside the main interferometer and compensate by increasing the input laser power}.

Thus the BHD photocurrent can be approximately expressed as:
\begin{multline}\label{eq:BHDphotos2}
I_{\rm BHD}\propto |\op{L}|(|e^0_0|\bigl[e^{2i\beta_{\rm arm}}(-|e^0_0|^2\mathcal{K}_{\rm MI}\hat{i}^0_c +\hat{i}^0_s)+\\
e^{i\beta_{\rm arm}}\dfrac{|e^0_0|\sqrt{2\mathcal{K}_{\rm MI}}}{x_{\rm SQL}}x_-\bigr]
 + \sqrt{1-|e^0_0|^2}\Delta\op{n}^{\rm HOM}_s)+h.c.
\end{multline}
So that the quantum noise power spectral density is given by:
\begin{equation}
S^{\rm QN} \simeq |e^0_0|^6 S^{0}_{s,\, BA}(f) + 1\,,
\end{equation}
and for the PSD of the QNLS, the above PSD is divided by the modulus squared of the optomechanical response function that is proportional to $|e^0_0|^4|\op{R}_{-s}|^2$:
\begin{equation}
S^{\rm QN}_x \simeq  \dfrac{|e^0_0|^2S^{0}_{s,\, BA}(f)}{|\op{R}_{-s}|^2} + \dfrac{1}{|e^0_0|^4|\op{R}_{-s}|^2}\,.
\end{equation}
So in back-action dominated frequency band the SNR is improved by $1/|e^0_0|^2$ due to lower power, circulating in the interferometer. While at the shot noise dominated band the SNR is decreased to a much stronger degree, \textit{i.e.} $|e^0_0|^4$, since the signal is reduced both, due to the misalignment of the LO beam, and due to the reduced response of the lower-power interferometer to the mirror displacement.

\subsection{Combined Output and Input Misalignment}

In Fig.~\ref{fig:InOutMisa}, we show a special case when input and output misalignment compensate each other so as to produce a perfect overlap of the LO and the signal beam at the BHD photodiodes. This somewhat artificial situation  demonstrates the fact that the effects of input and output misalignment can partially compensate each other. Here the reduction of SNR owes solely to the effect of the decrease of power circulating in the interferometer.

Hence, the QN PSD can be written as
\begin{equation}
S^{\rm QN} \simeq |e^0_0|^4 S^{0}_{s,\,BA}(f) + 1\,,
\end{equation}
and the response of the interferometer is reduced by the factor $|e^0_0|$. Combining these two effects in the QNLS PSD, one obtains: 

\begin{equation} 
S^{\rm QN}_x \simeq  \dfrac{|e^0_0|^2S^{0}_{s,\, BA}(f)}{|\op{R}_{-s}|^2} + \dfrac{1}{|e^0_0|^2|\op{R}_{-s}|^2}\,.
\end{equation}
For arbitrary misalignment combinations, the exact field distribution among different modes are necessary to be specified. In that case the general framework provided in Sec.~\ref{sec:qnbhd} can be used.

\section{Example of the Sagnac speed meter interferometer}
\label{sec:SSMdesign}

In this section, we give another illustrating example of the  influence of HOMs on the quantum noise, i.e. the  particular configuration of the zero-area Sagnac speed meter interferometer \cite{Beyersdorf99,chen2003} which is proposed as an candidate for supressing the SQL. To be specific, we use the parameters for the ERC-funded proof-of-principle prototype Sagnac speed meter interferometer (SSM) being constructed in the University of Glasgow \cite{G2014,Danilishin2015}, featuring  equivalent parameters as the Michelson configuration in the previous section.

We introduce general Sagnac interferometer with $R_{BS}$ and $T_{BS}$ representing the main BS power reflectivity and transmissivity.  
The interferometer transfer matrices, $\mathbb{A}$, $\mathbb{B}$, $\mathbb{C}$, $\mathbb{D}$, defined in Eqs.~\eqref{eq:I/O-rels_general} can be written for our particular case as \cite{Danilishin2015}:

\begin{subequations}
\begin{equation}
\mathbb{A}_{00}=
	2\sqrt{R_{BS}T_{BS}}e^{2i\beta_{\rm sag}}\begin{bmatrix}
	1 & 0 \\
	-\mathcal{K}_{\rm sym} & 1
	\end{bmatrix},
\end{equation}

\begin{equation}
\mathbb{B}_{00}=(R_{BS}-T_{BS})e^{2i\beta_{\rm sag}}
	\begin{bmatrix}
	1 & 0\\
	-4\mathcal{K}_{\rm MI} & 1
	\end{bmatrix},
\end{equation}

\begin{equation}
\mathbb{C}_{00}= (R_{BS}-T_{BS})e^{2i\beta_{\rm sag}}
	\begin{bmatrix}
	1 & 0\\
	0 & 1
	\end{bmatrix},
\end{equation}

\begin{equation}
\mathbb{D}_{00}= -2\sqrt{R_{BS}T_{BS}}e^{2i\beta_{\rm sag}}
	\begin{bmatrix}
	1 & 0\\
	-\mathcal{K}_{\rm asym} & 1
	\end{bmatrix}\,,
\end{equation}
\end{subequations}
where $\beta_{\rm sag}=2\beta_{\rm arm}+\frac{\pi}{2}$ is the corresponding phase shift for the full Sagnac interferometer.
The symmetric and asymmetric Saganac interferometer opto-mechanical coupling factors are defined as
\begin{equation}
\begin{split}
\mathcal{K}_{\rm sym} = 2\mathcal{K}_{\rm MI}\sin^2\beta_{\rm arm} \simeq \dfrac{8\Theta \gamma_{\rm arm}}{(\Omega^2+\gamma_{\rm arm}^2)^2}\,,
\\
\mathcal{K}_{\rm asym} = 2\mathcal{K}_{\rm MI}\cos^2\beta_{\rm arm} \simeq \dfrac{8\Theta \gamma^3_{\rm arm}}{\Omega^2(\Omega^2+\gamma_{\rm arm}^2)^2}\,,
\end{split}
\end{equation}

\begin{figure}
\centering
\includegraphics[width=1\columnwidth]{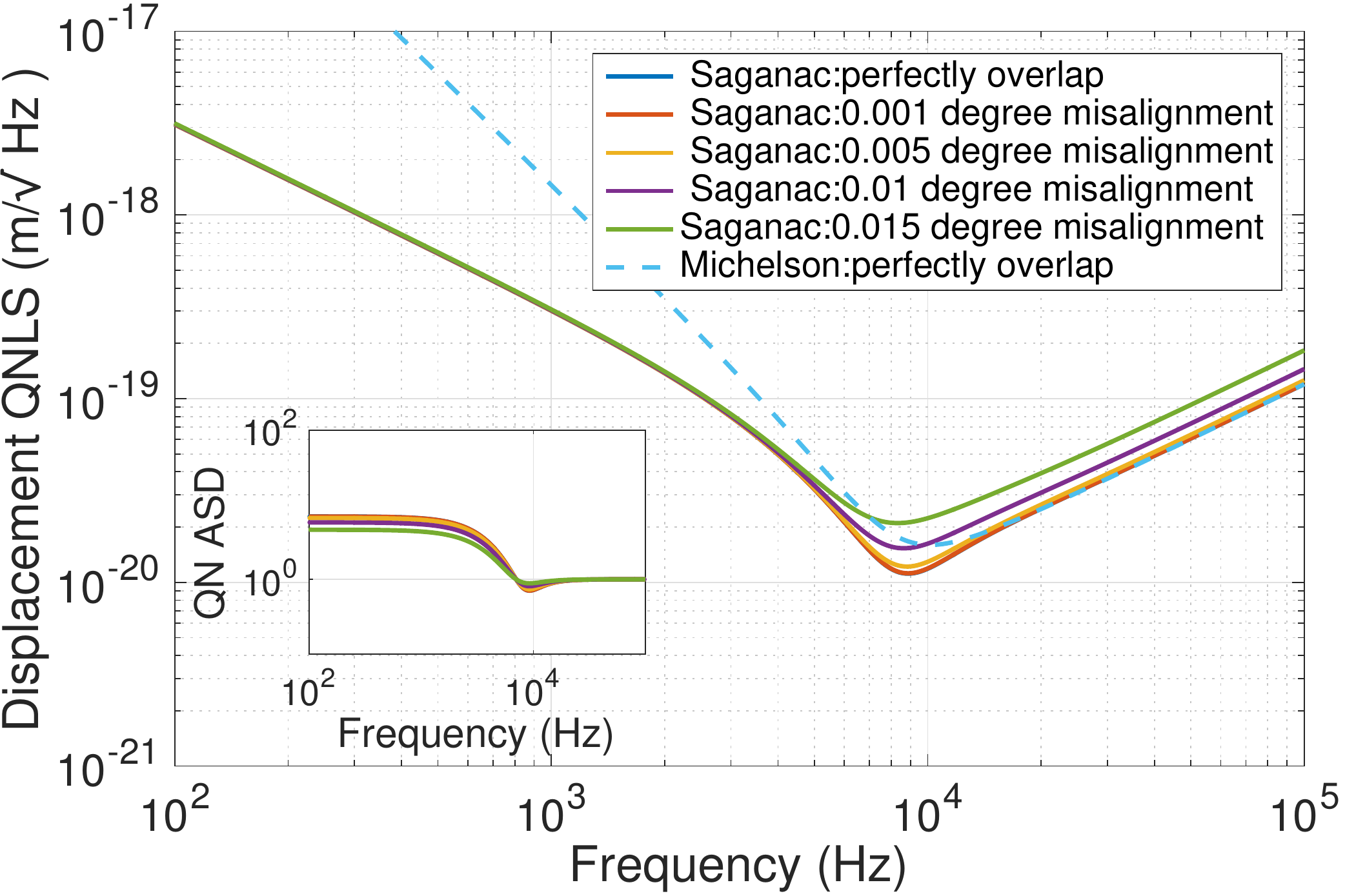}
\caption{Displacement quantum noise limited sensitivity (QNLS) of the Glasgow Sagnac interferometer for different values of misalignment angle between the LO beam and the signal one at the BHD. (The blue dashed curves indicates the QNLS of a perfectly aligned Michelson interferometer with equivalent parameters as the speedmeter.) Inset shows the amplitude spectral density of the QN only for the respective case. This gives the following values of equivalent lateral displacement of the two beams normalized by the beam radius on photodiode: 0.05, 0.25, 0.5, 0.7. } 
\label{fig:freeMisa}
\end{figure}

\begin{figure}
\centering
  \includegraphics[width=1\columnwidth]{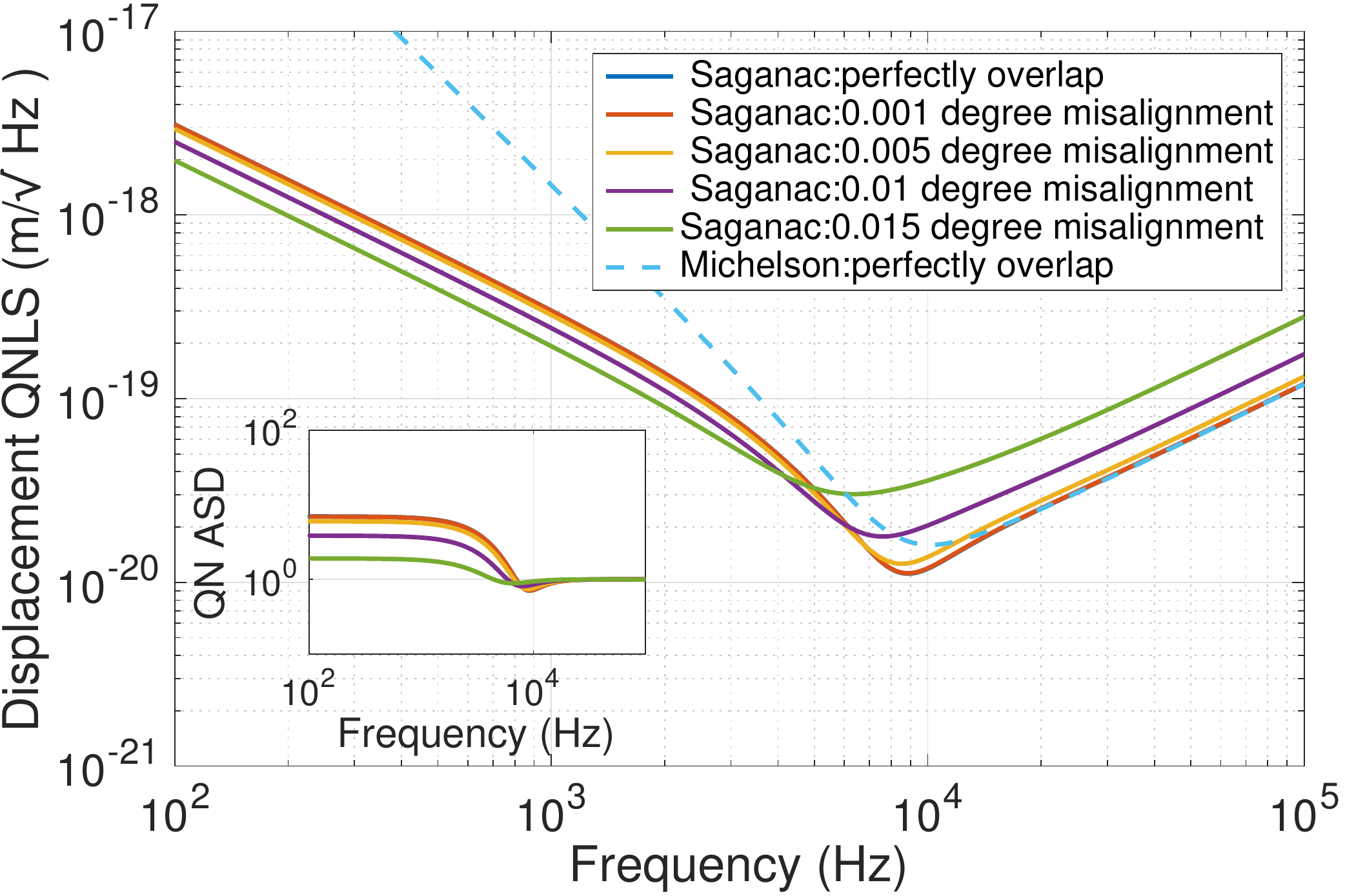}
\caption{ Displacement quantum noise limited sensitivity (QNLS) of the Glasgow Sagnac interferometer for different values of misalignment angle between input laser respect to the interferometer. (The blue dashed curves indicates the QNLS of a perfectly aligned Michelson interferometer with equivalent parameters asthe speedmeter.) Inset shows the amplitude spectral density of the QN only for the respective case. It is the same beam parameters and misalignment conditions as in Fig.~\ref{fig:freeMisa}. }
\label{fig:inputmis}
\end{figure}

The response of the interferometer to the common (cARM) and differential (dARM) mechanical modes of the arm mirrors, that are of particular interest in the context of gravitational wave detectors, can be written as:

\begin{eqnarray} 
&\pmb{R}_{\rm -} &= -ie^{2i\beta_{\rm arm}}\dfrac{\sqrt{2\mathcal{K}_{\rm sym}}}{x_{\rm SQL}}\begin{bmatrix}0\\1\end{bmatrix}\,,\\ 
&\pmb{R}_{\rm +} &= -e^{2i\beta_{\rm arm}}\dfrac{(R_{\rm BS}-T_{\rm BS})\sqrt{2\mathcal{K}_{\rm asym}}}{x_{\rm SQL}}\begin{bmatrix}0\\1\end{bmatrix}\,.
\end{eqnarray}
Fig.~\ref{fig:freeMisa} and Fig.~\ref{fig:inputmis} show the effect
of output and input misalignment of the Sagnac speedmeter with BHD readout, using similar levels of misalignment as were presented earlier for the example of the Fabry-P\'erot--Michelson interferometer. As expected, the observed effects from misalignment are the same for the Sagnac speedmeter and the Michelson interferometer.

\section{Non-stationary misalignment effects in the balanced homodyne detector}
\label{sec:jitter}

All the beams' misalignment and mismatch effects considered thus far were assumed stationary. However, in the real interferometer with suspended optics the optical paths of the LO and the signal beams are disturbed in a non-stationary way as a result of seismic motion of the ground. Therefore, the initially static parameters describing misalignment of Gaussian beams given in Sec.~\ref{sec:couplingc} now have to be considered as random functions of time defined by the local seismic noise of the lab. It is the subject of this section to estimate the additional noise in the BHD readout incurring from the random seismically driven movements of the suspended optical components, such as steering mirrors. Specifically, we look at the influence of tilt (pitch) motion, which has a much stronger coupling from the longitudinal ground motion than the rotation direction, which is a consequence of the suspension design. 

For simplicity, we assume the two input beams of the BHD, the LO and the signal beams, are Gaussian with non-zero DC components only in the fundamental TEM$_{00}$ mode, which can be justified by the use of output mode-cleaners for these two beams \cite{Fritschel2014}. We also assume AC parts, encompassing quantum and classical fluctuations, to be much smaller in magnitude than the DC components. 

The signal beam is the interformeter DP $\op o$, while the LO is taken form the reflected light $\op q$. As in Eq.~\eqref{eq:slfields}, we can thus write
\begin{equation}
\op L_0=\op Q_0,~ \op S_0=\mathbb{H}_{\phi_h}\mathbb{O}^{\textit{s-l}}_{00}(t) \op O_0,
\end{equation}
where $\op Q_0$ and $\op O_0$ are TEM$_{00}$ mode DC parts in $\op q$ and $\op o$, respectively. We choose the coordinate system of the LO beam as a reference, and the relative misalignment of the signal beam is represented by $\mathbb{O}_{00}^{\textit{s-l}}$, defined in terms of coupling coefficients $c_{0}^{0}$ as in Eq.~\eqref{eq:C-M}. According to Eq.~\eqref{eq:photocurrent}, the main dynamic photocurrents can be written as
\begin{equation}\label{eq:Idy}
I^{\rm dy}_{\rm BHD}\propto \op Q_0^{\dagger} \mathbb{H}_{\phi_h}\mathbb{O}^{\textit{s-l}}_{00}(t)\op O_0+h.c.
\end{equation}

We further assume that the two beams are perfectly matched in the static case, i.e. they have the same waist size $w_0$ and Rayleigh range $z_R$ and thus the same spot size on the photodetectors. According to Eq.~\eqref{eq:coeffs}, in misalignment condition $c_{0}^{0}(t)$ in terms of the small jitter angle $\theta$ or equivalent lateral beam shift $\Delta r$ and beam size $w(z)$ on the photodetectors is given by
\begin{multline}
c_{0}^{0}(t)=\exp\Bigl(-\frac{k^2 w^2(z)\Delta r^2(t)}{8(z^2+z_R^2)}\Bigr)
\\=\exp\Bigl(-\frac{k^2 w^2(z)\theta^2(t)}{8}\Bigr)\,,
\end{multline}
where $w(z)=w_0\sqrt{1+(z/z_R)^2}$. 
In general, $\theta(t)$ contains a DC part and the fluctuation part, which means $\theta(t)=\theta_{\rm DC} +\theta_{\rm fl}(t)$. 

In order to calculate the jitter noise spectral density, we need an additional step to calculate the spectral density of the quadratic random process ${\theta}^2(t)$, after which the spectral density of the jitter noise is straightforward to be written as
\begin{figure}[bt]
\centering
  \includegraphics[width=1\columnwidth]{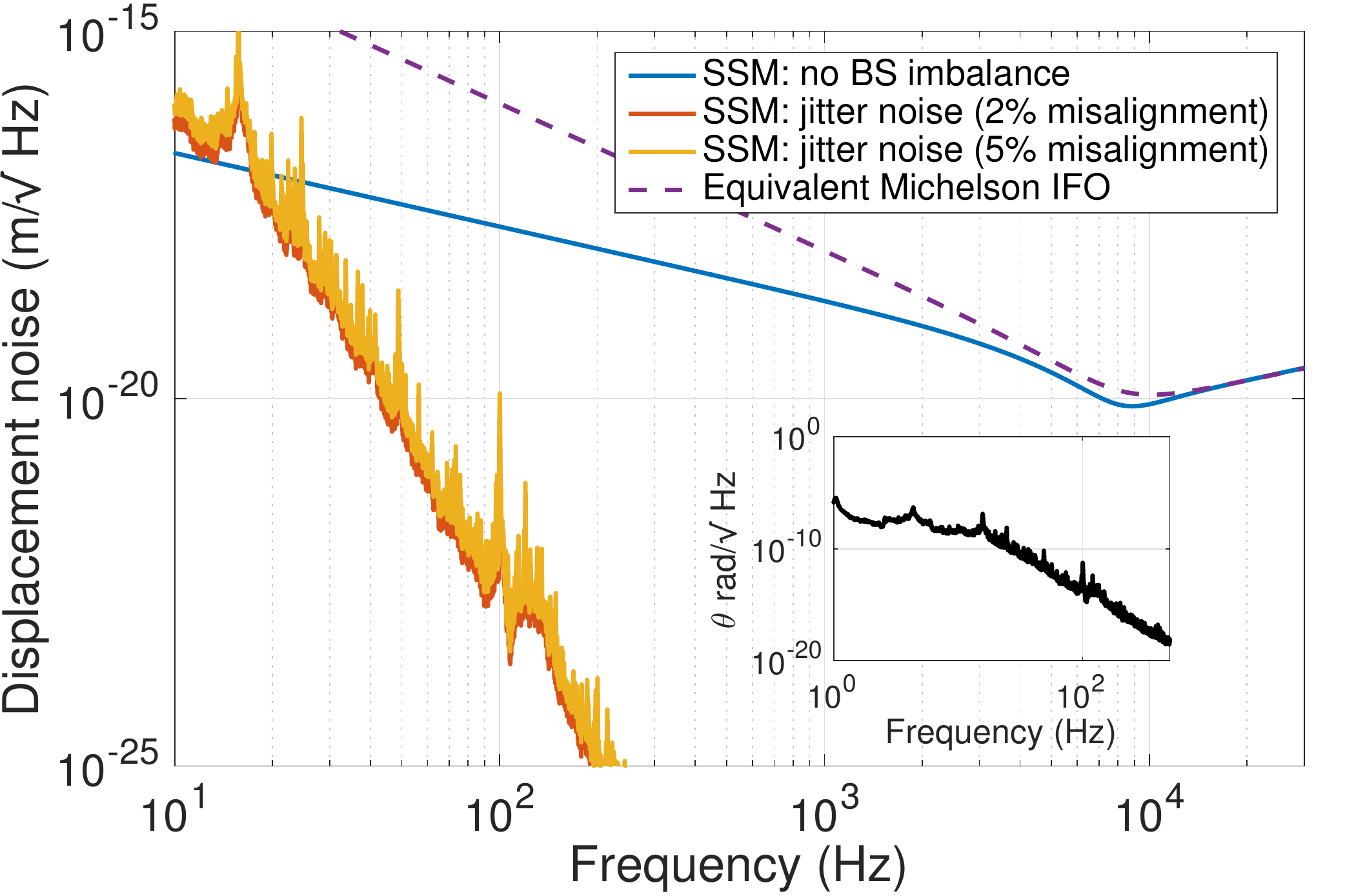}
\caption{ Additional non-stationary beam jitter noise driven by seismic motion in the lab plotted against QNLS curves of the equivalent Michelson interferometer and of the Sagnac speed meter interferometer with no imbalance in the beam-splitting ratio and with $0.1\%$ BS imbalance . 
The beam waist size is 0.925mm, with beam travel distance 2.038m. The spectral density of jitter angle $\theta$ is attached on the right lower corner.} 
\label{fig:cln}
\end{figure}
\begin{equation}\label{eq:Sjitter}
 S_{\rm jitter}=
\left|\op Q_0^{\dagger} \mathbb{H}_{\phi_h}\op O_0\right|^2 
\frac{k^4w^4(z)}{64} (4\theta^2_{\rm DC} S_{\theta_{\rm fl}} +S_{{\theta_{\rm fl}}^2}).
\end{equation}
According to Eq.~2-44 in \cite{ip1966},
$S_{\theta_{\rm fl}^2}$ turns out to be the convolution of the spectral density of $S_{\theta_{\rm fl}}(\Omega)$, which reads 
\begin{equation}\label{eq:len}
\begin{split}
S_{ \theta_{\rm fl} ^2}(\Omega)
=\int_{-\infty}^{\infty}S_{\theta_{\rm fl} }(\Omega') S_{\theta_{\rm fl} }(\Omega-\Omega')d\Omega,
\end{split}
\end{equation}
In Sagnac interferometer, the fundamental mode amplitude $\op O_0$ mainly comes from the imbalance of the main beamsplitter, $\eta_{BS}=R_{BS}-0.5$, then $\op O_0=2\eta_{BS}\op P_0$. 

Taking again the Glasgow SSM as an example, we calculated the expected additional non-stationary beam jitter noise due to sesmic motion coupling into the LO and signal path. While the double-pendulum suspensions of these mirrors \cite{G2014} strongly suppresses seismic noise at frequencies in our experiment band, there is still significant motion of the mirrors at the pendulum eigen frequencies. Starting from a measured displacement noise spectral density, we apply our simulated suspension transfer function for longitudinal motion to pitch motion coupling. This result in the pitch noise spectral density shown in the inset in Fig.~\ref{fig:cln}. From this, and using Eq.~\eqref{eq:Sjitter} and Eq.~\eqref{eq:len}, we arrive at the total noise contribution due to seismically driven beam-jitter noise in Glasgow SSM experiment as shown by the orange trace Fig.~\ref{fig:cln}.
This traditional noise is far below the quantum noise limited sensitivity in our measurement band between  $100-1000$ Hz. 

\section{summary}

In this article, we investigated the performance of balanced homodyne readout in practical applications including degradation effects from optical higher-order Hermite Gaussian modes. We provide a general solution for considering the effect of HOMs which are related to the input and output ports misalignment on the quantum noise limited sensitivity. It provides a framework for solving arbitrary conditions of input and output port misalignments or mismatch. This framework can be applied to any interferometer, i.e. it is independent of the actual interferometer configuration. We find that output port misalignments only degrades the amplitude spectral density of the shot noise limited part of the quantum noise noise sensitivity by a factor of $c_{0}^{0}$ or $d_{0}^{0}$, while the sensitivity in the back-action noise limited range will not degrade.  In the case of input misalignment, \textit{i.e.} the laser beam being misaligned in respect to the eigenmode of the interferometer, firstly the laser amplitude  inside the interferometer will be reduced by a factor $e_{0}^{0}$, thus changing the quantum noise limited sensitivity, and secondly it will also contribute to the LO beam misalignment and worsen the amplitude spectral density of the quantum noise limited sensitivity on high frequencies by a factor of ${e_{0}^{0}}^2$ in total. In addition, we investigated the noise coupling mechanisms from beam jitter, i.e. time varying HOM contributions.  Using the case  of the speed meter proof of concept experiment under construction in Glasgow as an illustrating example, we found that the seismically introduced beam jitter noise is well below the quantum noise level in our sensitive frequency range 10-1000\,Hz. We note that though our framework supports the injection of squeezed light states, for clarity we refrain from  a detailed discussion squeezing light injection in this article.

In conclusion, we have developed and applied a general framework for investigating realistic applications of balanced homodyne detection in suspended interferometers with realistic (\textit{i.e.} imperfect) optics, thus paving the way for technical design studies of future upgrades to gravitational wave detectors featuring balanced homodyne readout.  

\section{ackonwledgements}
The authors would like to thank colleagues from the
LIGO scientific collaboration, especially Keita Kawabe for fruitful discussion.
The work described in this article is funded by the European
Research Council (ERC-2012-StG: 307245).  We are
grateful for support from the Science and Technology
Facilities Council (Grant No. ST/L000946/1). SS was supported by the European Commission H2020-MSCA-IF2014 actions, grant agreement number 658366.
\bibliography{bhd_mismatch_v5}
\end{document}